\documentclass[aps,amsmath,amssymb,preprintnumbers]{revtex4}
\usepackage{graphicx}

\begin{document}

\title{Geometric approach to asymptotic expansion of Feynman integrals}
\author{A. Pak$^{a}$ and A. Smirnov$^{a,b}$}
\affiliation{
  $^{a}$ Institut f{\"u}r Theoretische Teilchenphysik,
  Karlsruhe Institute of Technology (KIT),
  76128 Karlsruhe, Germany \\
  $^{b}$ Scientific Research Computing Center,
  Moscow State University,
  119991 Moscow, Russia}
\begin{abstract}
We present an algorithm that reveals relevant contributions
in non-threshold-type asymptotic expansion of Feynman integrals
about a small parameter. It is shown that the problem reduces
to finding a convex hull of a set of points in a
multidimensional vector space.
\end{abstract}
\maketitle

\section{Introduction}

Evaluation of Feynman integrals depending on multiple parameters
is a notoriously difficult task. When direct computation fails,
one resorts to studying asymptotics in various limits. In practice,
a few first terms in the expansion may already suffice to 
reach the desired precision.
However, expansion of a multi-loop integral may become non-trivial
due to an interplay of parameters with the integration variables
(components of loop momenta). Classification of relevant sectors
in the integration space is itself a challenging
problem~\cite{Beneke:1997zp,Smirnov:1999bza}.

One important case is the asymptotic expansion in momenta and
masses in the limits typical for the Euclidean space. This problem has
been completely solved in terms of sums over subgraphs~\cite{Chetyrkin:1988zz,
Chetyrkin:1988cu,Tkachov:1997gz,Gorishny:1989,Smirnov:1990rz}.
At least one automated tool~\cite{Seidensticker:1999bb,Harlander:1997zb}
implements this approach in practice.
For a more general situation, including the limits appearing in
the Minkowski space, there exists the universal strategy of
expansion by regions in the momentum space
~\cite{Beneke:1997zp,Smirnov:1999bza,Smirnov:2002pj}.
In all known cases, it produces correct results,
but a rigorous proof is still lacking. Typically, one manually
analyzes a multi-scale problem, starting from simpler examples
that can be checked against known analytical results
or numerical estimates, computed e.g. with {\tt FIESTA}~\cite{Smirnov:2008py}
(later versions~\cite{Smirnov:2009pb} of FIESTA may also evaluate
a few first terms in a given asymptotic expansion.)

An important type of non-Euclidean expansions, the so-called threshold
expansion~\cite{Beneke:1997zp}, requires the most careful treatment.
Cancellation of dominant terms becomes obvious only in a specially
chosen frame or with a certain routing of loop momenta.
In what follows we try to elaborate some approach to non-threshold
asymptotic expansion, based on alpha-representation of integrals, and
describe a simple practical algorithm.

\section{Expansion by regions and alpha-representation}

A thorough introduction to the expansion by regions and alpha-representation
can be found elsewhere~\cite{Smirnov:2002pj,Bogolyubov:1980nc}. Here we
briefly introduce the basic notation with a trivial example.

Consider a family of one-loop propagator-type integrals in the Euclidean space:
\begin{eqnarray}
  I_1(a_1,a_2;p^2,m^2) &=& \int\frac{d^D k}{(2\pi)^D}
    \frac{1}{D_1^{a_1} D_2^{a_2}},~~
  D_1 = k^2 + m^2, ~~D_2 = (k+p)^2 + m^2.
  \label{eqn:1e1l2s}
\end{eqnarray}
A specific integral is determined by the exponents $a_1$ and $a_2$
and depends on the two parameters, $m^2$ and $p^2$.
The structure of the expansion does not depend on
$a_1$ and $a_2$ and we will not mention those exponents in the
following discussion.

We consider the asymptotics of
$I_1(a_1,a_2;p^2,m^2)$ in the limit when $|p^2| \gg m^2$,
or $\rho = |m^2/p^2| \ll 1$. The naive Taylor expansion does not 
capture the complete asymptotic behaviour since the integration
variables (components of $k$) span all values from $-\infty$ to
$+\infty$, and in particular can be as small as $m$ or as large 
as $\sqrt{|p^2|}$.

\begin{table}[t]
  \begin{tabular}{|c|l|l|l|}
  \hline
  \parbox[l]{0.25\textwidth}{\vspace{0.1cm}
    \includegraphics[width=0.2\textwidth]{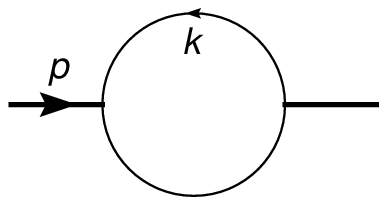}
    \put(-15,40){(a)} \\ \vspace{0.1cm}} &
  \parbox[l]{0.3\textwidth}{
    \begin{eqnarray*}
      D_1 &=& k^2 + m^2, \\
      D_2 &=& (k+p)^2 + m^2, \\
      \rho &=& |m^2/p^2| \ll 1
    \end{eqnarray*}} &
  \parbox[l]{0.4\textwidth}{
    \begin{eqnarray*}
      \mathcal{U} &=& x_1 + x_2, \\
      \mathcal{F} &=& x_1 x_2 (p^2 + 2 m^2) + x_1^2 m^2 + x_2^2 m^2
    \end{eqnarray*}} \\ \hline
  \parbox[l]{0.25\textwidth}{\vspace{0.1cm}
    \includegraphics[width=0.2\textwidth]{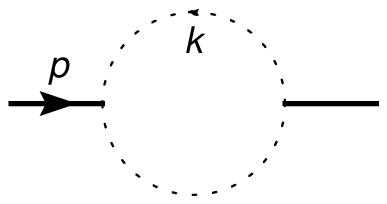}
    \put(-15,40){(b)} \\ \vspace{0.1cm}} &
  \parbox[l]{0.3\textwidth}{
    $|k^2| \sim |p^2| \gg m^2$,
    \begin{eqnarray*}
      D_1^{(b)} &=& k^2,~~
      D_2^{(b)} = (k+p)^2
    \end{eqnarray*}} &
  \parbox[l]{0.4\textwidth}{
    $x_1,x_2 \sim \rho^A$,
    \begin{eqnarray*}
      \mathcal{U}^{(b)} &=& x_1 + x_2,~~
      \mathcal{F}^{(b)} = x_1 x_2
    \end{eqnarray*}} \\ \hline
  \parbox[l]{0.25\textwidth}{\vspace{0.1cm}
    \includegraphics[width=0.2\textwidth]{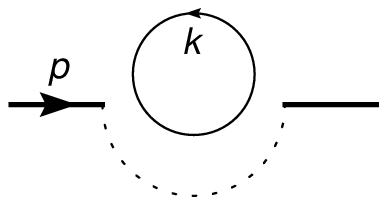}
    \put(-15,40){(c)} \\ \vspace{0.1cm}} &
  \parbox[l]{0.3\textwidth}{
    $|k^2|\sim m^2$,
    \begin{eqnarray*}
      D_1^{(c)} &=& k^2 + m^2,~~
      D_2^{(c)} = p^2
    \end{eqnarray*}} &
  \parbox[l]{0.4\textwidth}{
    $x_2\sim \rho^A$, $x_1\sim \rho^{A-1}$,
    \begin{eqnarray*}
      \mathcal{U}^{(c)} &=& x_1,~~
      \mathcal{F}^{(c)} = x_1 x_2 p^2 + x_1^2 m^2
    \end{eqnarray*}} \\ \hline
  \parbox[l]{0.25\textwidth}{\vspace{0.1cm}
    \includegraphics[width=0.2\textwidth]{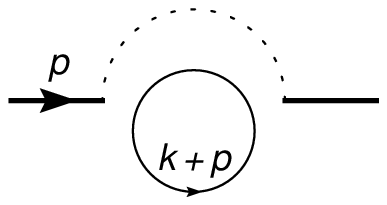}
    \put(-15,40){(d)} \\ \vspace{0.1cm}} &
  \parbox[l]{0.3\textwidth}{
    $|(k+p)^2|\sim m^2$,
    \begin{eqnarray*}
      D_1^{(d)} &=& p^2,~~
      D_2^{(d)} = (k+p)^2 + m^2
    \end{eqnarray*}} &
  \parbox[l]{0.4\textwidth}{
    $x_2\sim\rho^{A-1}$, $x_1\sim\rho^A$
    \begin{eqnarray*}
      \mathcal{U}^{(d)} &=& x_2,~~
      \mathcal{F}^{(d)} = x_1 x_2 p^2 + x_2^2 m^2
    \end{eqnarray*}} \\ \hline
  \end{tabular}
  \caption{\label{tab:ex1}Regions (b-d) of expansion 
    of a double-scale integral (a) in the momentum
    space and in the alpha-representation.}
\end{table}

The prescription in this case is to find regions, or
scalings of momentum components that after the expansion
provide non-zero contributions. In each region, we first
Taylor expand the integrand and drop the scaling restrictions.
In our example, there are three non-zero regions, summarized
in Tab.~\ref{tab:ex1}. For example, in the region (c)
one expands $D_2$ as follows:
\begin{eqnarray}
  I_1(a_1,a_2;p^2,m^2) &=&
    \int_{k_\mu \sim m} \frac{d^D k}{(2\pi)^D}
    \frac{1}{(k^2 + m^2)^{a_1}}
    \left[\frac{1}{(p^2)^{a_2}} - \frac{a_2(k^2 + 2kp + m^2)}{(p^2)^{a_2 + 1}}
      + \ldots\right] + \mbox{other regions}
    \\ \nonumber
    &=&
    \left[
    \int\frac{d^D k}{(2\pi)^D}
    \frac{1}{(k^2 + m^2)^{a_1}(p^2)^{a_2}}
    +
    \int\frac{d^D k}{(2\pi)^D}
    \frac{(...)}{(k^2 + m^2)^{a_1}(p^2)^{a_2+1}} + \ldots
    \right]
      + \mbox{other regions}.
\end{eqnarray}
In the last line, we dropped the restriction on $k$ and the problem
reduced to evaluation of (multiple) Feynman integrals with simpler
denominator factors, various denominator exponents and possibly more complex
numerators. The non-trivial statement is that the double-counting which
could have been introduced disappears in the sum of all regions. For the
purpose of the following discussion we assume that the tensor reduction
of numerators and the proliferation of terms can be managed; we will
focus on the transformations of denominator factors in every region
(e.g. $(k + p)^2 + m^2 \to p^2$ in the example above).

Some types of asymptotic expansion may require more elaborate choice of
regions. For example, $k_0$ may scale differently from $k_i$ or
some combination of components may have a separate scale (this
happens, e.g., in Sudakov limits).
It is thus desirable to have an explicitly covariant formalism to
identify regions, independent of the frame choice and the
routing of momenta. For that purpose, we may switch to the
alpha-representation of integrals. We re-write an integral with
$n$ lines (denominator factors) over $D$-dimensional loop momenta
as an integral over $n$ positive parameters $x_1,..,x_n$.
Information about the graph is then encoded in the two homogeneous
polynomials, $\mathcal{U}$ and $\mathcal{F}$. For example, our
integral above is
\begin{eqnarray}
  I_1(a_1,a_2;p^2,m^2) &=& 
    \frac{\Gamma(a_1 + a_2 - D/2)}{\Gamma(a_1)\Gamma(a_2)}
    \int_0^\infty dx_1 dx_2 \delta(1 - x_1 - x_2) x_1^{a_1-1} x_2^{a_2-1}
    \mathcal{U}^{a_1+a_2-D}\mathcal{F}^{D/2-a_1-a_2}, \\ \nonumber
    && 
    \mathcal{U} = x_1 + x_2,~~
    \mathcal{F} = x_1 x_2 (p^2 + 2 m^2) + x_1^2 m^2 + x_2^2 m^2.
  \label{eqn:1e1l2a}
\end{eqnarray}

The expansion by regions may also be formulated in the alpha-representation
~\cite{Smirnov:1999bza,Smirnov:2002pj}. Instead of finding the scaling
behaviour of loop momentum components, here we deal with the scaling
of each parameter $x_i$ that directly corresponds to the scale of the
$i$-th line (denominator factor) of the original integral.
During the expansion, only the leading terms remain in the polynomials
$\mathcal{U}$ and $\mathcal{F}$, and the resulting alpha-representation
represents the integrals obtained by the expansion in the momentum space.
The last column of Tab.~\ref{tab:ex1} demonstrates the scaling of
alpha-parameters and the polynomials corresponding to each region.

Note that in the language of alpha-parameters, the difference between
the threshold-type and non-threshold expansion becomes clear.
Let us consider integral of Eq.~\ref{eqn:1e1l2s}
in the threshold limit $y = m^2 + \frac{p^2}{4} \ll m^2$ (that, of course,
implies that $p^2 < 0$, i.e. this limit is essentially non-Euclidean).
Choosing the frame where $p = (p_0,\vec{0})$ and re-routing the loop
momentum, we obtain the denominator factors $D_1 = k_0^2 + \vec{k}^2 + k_0 p_0 + y$
and $D_2 = k_0^2 + \vec{k}^2 - k_0 p_0 + y$.

This integral has two
non-vanishing regions. The first ``hard'' region is characterized by
$k\sim m$. The second ``potential'' region corresponds to
$k_0\sim y/m$, $|\vec{k}|\sim \sqrt{y}$. In the language of
alpha-parameters, in the ``hard'' region only the
second term survives in the polynomial
$\mathcal{F} = y(x_1 + x_2)^2 - \frac{p^2}{4} (x_1 - x_2)^2$.
The most troublesome ``potential'' region stems from a thin layer
in the integration space near the surface $x_1 = x_2$, when
the second term has the same scaling as the first.

In a similar way, more complex threshold expansions receive
contributions which depend on cancellations between the terms in
the expanded $\mathcal{F}$ which happens along some
non-trivial surface and not at zero or infinity. Presently we do
not know a general rule to identify such surfaces and find
substitutions revealing such regions. Instead, we focus on the
``usual'' regions that can be determined by examining
independently the monomials in $\mathcal{U}$ and $\mathcal{F}$.
However limited, this problem is still important for many applications.

\section{General formalism}

We consider an $l$-loop Feynman integral
\begin{eqnarray}
  I(a_1,...,a_n) &=& \int
    \frac{d^D k_1...d^D k_l }{(2\pi)^{lD} D_1^{a_1}...D_n^{a_n}},
  \label{eqn:momrep}
\end{eqnarray}
which depends on $n$ exponents $a_1,...,a_n$, scalar products of $e$
external momenta $p_1,...,p_e$ and parameters (such as masses) in
denominator factors $D_i$. The latter must be quadratic in momenta but
other than that may have any form, e.g. correspond to a massive,
such as $-(k_i + p_j)^2 + m_k^2 - i0$, or a static propagator,
such as $(- 2 k_i p_j \pm i0)$.
The alpha-representation for this integral has a general structure
\begin{eqnarray}
  I(a_1,...,a_n) &=& c \int_0^1 d x_1 ... d x_n~
  \delta(1-x_1-...-x_n) x_1^{a_1-1}...x_n^{a_n-1}
  \mathcal{U}^a\mathcal{F}^b,
  \label{eqn:alprep}
\end{eqnarray}
where coefficient $c$ and exponents $a$ and $b$ depend only on $l$,
$D$, and $a_i$. $\mathcal{U}$ and $\mathcal{F}$ are homogeneous
polynomials (of order $l$ and $l+1$, respectively) of integration
variables $x_i$, and $\mathcal{F}$ also depends on the kinematic invariants.
If the denominators $D_i$ correspond to some graph and have a standard 
form $- k^2 + m^2 - i0$ (in Minkowski space), then the functions
$\mathcal{U}$ and $\mathcal{F}$ can be read off the graph in terms of
trees and 2-trees~\cite{Bogolyubov:1980nc}. In a more general case,
one may obtain $\mathcal{U}$ and $\mathcal{F}$ with a tool found at
\verb+http://www-ttp.particle.uni-karlsruhe.de/~asmirnov/Tools-UF.htm+.
In what follows, we will only discuss the properties of $\mathcal{U}$ and
$\mathcal{F}$ that are independent of specific indices $a_i$.

In dimensional regularization, ``scaleless'' integrals (having no inherent
scale) turn to zero. More specifically, an integral is scaleless if it is
possible to re-scale some loop momenta or their components so that the result
remains proportional to the original integral, or
$D_i(\{k_j\},\{a k_i\}) = a^{u_i} D_i(\{k\})$,
with some subset $\{k_i\}$ of integration momenta. In particular,
massless vacuum bubbles vanish:
\begin{equation}
  I = \int\frac{d^D k}{(k^2)^n}
    = \int\frac{d^D (\alpha k)}{((\alpha k)^2)^n}
    = \alpha^{D-2n} I = 0.
\end{equation}

In terms of the alpha-representation Eq.~\ref{eqn:alprep}, a similar statement
applies to homogeneity of $\mathcal{U}$ and $\mathcal{F}$ with respect to
a subset $\{B\}$ of integration variables $x_i$ ($\{B\}$ should not coincide
with the full set of $\{x_i\}$). Integrals vanish if
$\mathcal{U}(\{x_j\},\{a x_i\}) = a^u \mathcal{U}(\{x\})$ and
$\mathcal{F}(\{x_j\},\{a x_i\}) = a^f \mathcal{F}(\{x\})$, $i\in \{B\}$,
with some scaling dimensions $u$ and $f$.

In order to avoid separate treatment of $\mathcal{U}$ and $\mathcal{F}$,
one may consider the product $\mathcal{UF}$ that incorporates the scaling
and asymptotic properties of both factors (but may contain many terms).

\section{Geometric interpretation of asymptotic expansion}

Let us start with some integral in the alpha-representation Eq.~\ref{eqn:alprep}
with integration variables $x_1,...,x_n$ and a small expansion parameter $\rho$.
Each of $M$ terms in $\mathcal{F}$ corresponds to a
vector of $n+1$ exponents (we here neglect common factors and
numeric coefficients, irrelevant to the non-threshold expansion):
\begin{equation}
  \rho^{r_0} x_1^{r_1}...x_n^{r_n}\to (r_0,r_1,...,r_n),
\end{equation}
and $\mathcal{F}$ corresponds to a set $\{F\}$ of $M$ points
in $(n+1)$-dimensional vector space. Due to homogeneity of $\mathcal{F}$,
all these points belong to an $n$-dimensional hyperplane $r_1+...+r_n = l+1$,
parallel to the 0-th axis (the axis of $r_0$).

\begin{figure}[t]
  \centering
  \includegraphics[width=0.5\textwidth]{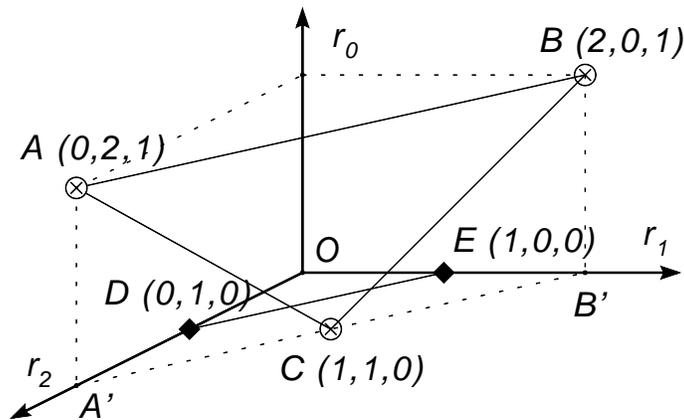}
  \caption[]{\label{fig:geom}
    Graphical representation of sets $\{F\}$ (crossed points)
    and $\{U\}$ (diamonds) corresponding to the integral
    of Eq.~\ref{eqn:1e1l2s}.}
\end{figure}

Terms of $\mathcal{U}$ have no explicit powers of $\rho$ in the
coefficients. The corresponding set $\{U\}$ is thus confined to
an $(n-2)$-dimensional hyperplane $r_0 = 0$, $r_1+...+r_n = l$.
In Fig.~\ref{fig:geom} we present such points corresponding to
the example in Eq.~\ref{eqn:1e1l2s}, where the three terms of
$\{F\}$ are denoted with crossed points and the two terms of 
$\{U\}$ with diamonds.

If we fix the scales of alpha-parameters as $x_i\sim \rho^{v_i}$, then
the scale of a monomial is
$\rho^{r_0} x_1^{r_1}...x_n^{r_n} \sim
 \rho^{r_0 + v_1 r_1 + ... + r_n v_n} \sim \rho^{\vec{r}\vec{v}}$
with $\vec{r} = (r_0,...,r_n)$ from $\{F\}$ and $\vec{v} = (1,v_1,...,v_n)$.
Graphically, $\vec{r}\vec{v}$ represents the length of a projection
of the vector $\vec{r}$ on the direction $\vec{v}$.

Some special choices of directions $\vec{v}$ determine the
regions of expansion that we seek. The terms in $\mathcal{F}$ that remain
after the expansion are all characterized by the same scale in powers of $\rho$.
All points of the corresponding subset $\{F^\prime\}$ then feature the same
value of the projection on $\vec{v}$, i.e. these points belong to the
hyperplane orthogonal to $\vec{v}$.

The points corresponding to the neglected
terms will be located ``above'' this hyperplane (since
$\vec{v}$ always points ``up'' with respect to the 0-th axis). In other words,
$\{F^\prime\}$ belong to a facet of the envelope, or the ``convex hull'' of
the set $\{F\}$, while the corresponding $\vec{v}$ is the normal vector to
that facet. In a similar manner we may define a subset
$\{U^\prime\}$ of the remaining terms in $\mathcal{U}$.

Relating the three expansion regions of Tab.~\ref{tab:ex1} to
graphics in Fig.~\ref{fig:geom}, we find the corresponding 
points and vectors (points as denoted in the figure):
\begin{itemize}
\item (b): $\vec{v} = (1,0,0)$, $\{F^\prime\} = (C)$, $\{U^\prime\} = (D,E)$,
\item (c): $\vec{v} = (1,1,-1)$, $\{F^\prime\} = (A,C)$, $\{U^\prime\} = (D)$,
\item (d): $\vec{v} = (1,-1,1)$, $\{F^\prime\} = (B,C)$, $\{U^\prime\} = (E)$.
\end{itemize}

Here we exploit the freedom to re-scale all $x_i$ by the same power of $\rho$,
i.e. shift $\vec{v}$ by any vector $\vec{a} = (0,A,...,A)$. If $\vec{v}$ 
corresponds to a region, then $\vec{v^\prime} = \vec{v} + \vec{a}$ determines
the same region. For example, $\vec{v^\prime} = (1,2,0)$ also corresponds to
the region (c) above. It is convenient to choose $\vec{v}$
parallel to the plane where points $\{F\}$ are confined, i.e. orthogonal to
the vector $(0,1,...,1)$.

Normally, only a few scaling choices produce non-zero regions. In our example,
the choice $x_1\sim \rho^2$, $x_2\sim\rho^0$, leading to $\mathcal{U} = x_2$,
$\mathcal{F} = x_2^2 m^2$, or $\{F^\prime\} = (A)$, $\{U^\prime\} = (D)$,
corresponds to a scaleless integral. As discussed above, this implies an
existence of a scaling leaving both $\mathcal{U}$ and $\mathcal{F}$ invariant
up to a pre-factor (in this case, $x_1\to a x_1$).

The requirement that a region does not vanish can be easily formulated in the
geometrical language. Consider the polynomial $\mathcal{UF}$ and the
corresponding set of points $\{UF\}$. After the expansion with the chosen
scalings, we are left with its subset $\{UF^{\prime}\}$. Numeric
coefficients and kinematic invariants are irrelevant to the scalefulness of the
region, and we get rid of them by projecting $\{UF^{\prime}\}$
on the plane $r_0 = 0$. The thus obtained set of points
$\{UF^{\prime}_0\}$ belongs to the $(n-1)$-dimensional hyperplane.

Scalelessness implies that all terms of the polynomial $\mathcal{UF}^\prime$
are homogeneous with respect to a certain re-scaling. We thus deduce
that the points of $\{UF^{\prime}_0\}$ must then
belong to an orthogonal space of the corresponding
re-scaling vector $\vec{v}_h$. In other words, $\{UF^{\prime}_0\}$
is confined to at most $(n-2)$-dimensional subspace, and
its $(n-1)$-dimensional volume is zero. The latter property
can be easily checked (and used to check whether a given 
integral vanishes). However, it is easy to see that the
``bottom'' facets of the convex hull for $\{UF\}$ 
automatically correspond to scaleful regions: their dimension
is $(n-1)$ by construction (otherwise they become ``ridges'' or
``vertices''), and they (by selection) are not orthogonal
to the plane $r_0 = 0$ (thus the projection has non-zero volume). 

Finally, we may formulate the general procedure to determine the
expansion regions. We start by building the set of points $\{UF\}$.
Next, we find the $n$-dimensional convex hull $\mathcal{C}$ of
the set $\{UF\}$ in the $n$-dimensional plane $r_1+...+r_n = l+1$,
using any preferred algorithm.
The implementation that we chose, QHull~\cite{QHull}, does not allow building
hulls of dimensionality lower than the dimension of vector space.
Thus, one has to introduce local coordinate system and deal
with non-integer coordinate values.
However, it is also possible to project $\{UF\}$ along any of
axes $r_i$, $i\ne 0$, e.g. consider $(n-1)$-dimensional
points $\vec{r}_{\parallel} = (r_0,r_1,...,r_{n-1})$.
Convex hull $\mathcal{C}^\prime$ built for this projection will be
the projection of the ``true'' convex hull $\mathcal{C}$.
Its dimensions will be stretched but the correspondence of the
points to the facets and the vertices will persist.

From the $(n-1)$-dimensional facets of $\mathcal{C}$ we then select
the ``bottom'', i.e. facets with normal vectors $\vec{v}$
pointing ``up'', with $v_0 > 0$.
For each of those ``bottom'' facets, we choose the normal vector
$\vec{v}$ such that $v_0 = 1$. Its components 1 to $n$
represent the relative scales of alpha-parameters $x_i$
and thus uniquely determine an expansion region.

\section{\label{sec:lte}Less trivial example}

\begin{figure}[t]
  \centering
  \includegraphics[width=0.2\textwidth]{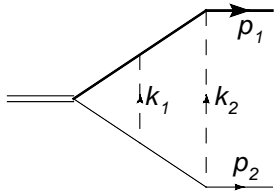}
  \caption[]{\label{fig:3e2l2s}
    Double-scale two-loop vertex integral.}
\end{figure}

Let us consider the integral in Fig.~\ref{fig:3e2l2s}, this
time defined in the Minkowski space (this example was first
considered in~\cite{Smirnov:1999bza} and~\cite{Smirnov:2002pj},
Chapter 10):
\begin{eqnarray}
  && I_2(a_1,...,a_6; s,m^2,M^2) = \int
    \frac{d^D k_1 d^D k_2}
    {(2\pi)^{2D} D_1^{a_1} ... D_6^{a_6}},
  \\ \nonumber &&
  D_1 = (p_1 - k_1 - k_2)^2 - M^2,~~
  D_2 = (p_1 - k_2)^2 - M^2,~~
  D_3 = (p_2 + k_1 + k_2)^2 - m^2,~~
  \\ \nonumber &&
  D_4 = (p_2 + k_2)^2 - m^2,~~
  D_5 = k_1^2,~~
  D_6 = k_2^2,~~
  \\ \nonumber &&
  p_1^2 = M^2,~~
  p_2^2 = m^2,~~
  (p_1 + p_2)^2 = s,~~
  s,M^2 \gg m^2.
  \label{eqn:3e2l2s}
\end{eqnarray}

With $S = m^2 + M^2 - s = - 2 p_1 p_2$, its alpha-representation reads:
\begin{eqnarray}
  \label{eqn:fex}
  \mathcal{U} &=& x_1 x_2 + x_3 x_2 + x_5 x_2
    + x_1 x_4 + x_3 x_4 + x_1 x_5 + x_3 x_5
    + x_4 x_5 + x_1 x_6 + x_3 x_6 + x_5 x_6,
  \\ \nonumber 
  \mathcal{F} &=& 
      M^2 x_1^2 x_2
    + M^2 x_1^2 x_4
    + M^2 x_1^2 x_5
    + M^2 x_1 x_2^2
    + M^2 x_2^2 x_3 
    + M^2 x_2^2 x_5 
    + M^2 x_1^2 x_6
    + 2 M^2 x_1 x_2 x_5
    \\ \nonumber &+&
      m^2 x_2 x_3^2
    + m^2 x_3^2 x_4 
    + m^2 x_1 x_4^2
    + m^2 x_3 x_4^2
    + m^2 x_3^2 x_5
    + m^2 x_4^2 x_5
    + m^2 x_3^2 x_6 
    + 2 m^2 x_3 x_4 x_5
    \\ \nonumber &+&
      S x_1 x_2 x_3
    + S x_1 x_2 x_4
    + S x_1 x_3 x_4
    + S x_1 x_3 x_5
    + S x_1 x_4 x_5
    + S x_1 x_3 x_6 
    + S x_2 x_3 x_4 
    + S x_2 x_3 x_5 
    + S x_2 x_4 x_5. 
\end{eqnarray}

For simplicity, let us analyze $\mathcal{F}$
instead of the product $\mathcal{UF}$. Choosing $m$ as the
small parameter and preserving the order of terms, in
the 7-dimensional space we have 25 points:
$\{F\}$ = (0,2,1,0,0,0,0), (0,2,0,0,1,0,0), (0,2,0,0,0,1,0), \ldots.

The projection of $\{F\}$ along the 6-th axis (i.e. 
$\{F\}_p$ = (0,2,1,0,0,0), (0,2,0,0,1,0), \ldots)
has a six-dimensional convex hull with 18 facets.
Of them, four belong to the ``bottom''. Restoring the
7-dimensional normal vectors with unit 0-th component, we find:
$\vec{v}_1 = (1,0,-2,-2,-2,-2,-4)$,
$\vec{v}_2 = (1,0,0,-2,-2,-2,-2)$,
$\vec{v}_3 = (1,0,0,-2,0,0,0)$,
$\vec{v}_4 = (1,0,0,0,-2,0,-2)$.
Since $\mathcal{U}$ is scaleful, we also have to add the ``hard''
region: $\vec{v}_0 = (1,0,0,0,0,0,0)$. (The latter would appear
automatically, had we used the product $\mathcal{UF}$.)

For illustration, let us consider the most non-trivial 
``ultrasoft-collinear'' region corresponding to $\vec{v}_1$,
or the scaling of alpha-parameters
$x_1\sim m^0$,
$x_2\sim 1/m^2$,
$x_3\sim 1/m^2$,
$x_4\sim 1/m^2$,
$x_5\sim 1/m^2$, and
$x_6\sim 1/m^4$.
First, we may check that $\vec{v}_1$ is indeed orthogonal to
the plane containing the points 5, 6, 15, 22, 23, 24, and 25
from $\{F\}$ (in the order as in Eq.~\ref{eqn:fex}).
Those points correspond to the terms remaining in $\mathcal{F}^\prime$
after the expansion: $M^2 x_2^2 x_3 + M^2 x_2^2 x_5 + \ldots$.

In the momentum space, the interpretation becomes clear only
in the special reference frame, where
$p_1 = (M,\vec{0},0)$,
$p_2 = M n_+ + \left(\frac{m^2}{M}\right) n_-$,
and
$n_{\pm} = (1/2,\vec{0},\mp 1/2)$.
We also decompose the first loop momentum in plus- and
minus- parts, $k_1 = (k_+ + k_-,\vec{k},k_+ - k_-)$.
To reproduce the ``ultrasoft-collinear'' region, we should
prescribe the following scales to the components of loop momenta:
$k_+\sim m^2/M$, $k_-\sim M$, $\vec{k}\sim m$, $k_2\sim m^2/M$.

After the expansion, the denominator factors scale as:
$D_1\sim M^2$, $D_2,D_3,D_4,D_5\sim m^2$, $D^6\sim m^4/M^2$.
One can easily see how the powers of $m$ here correspond
to the components of $-\vec{v}_1$.

\section{Implementation}

We wrote a Mathematica program that determines the
expansion regions of a given Feynman integral based on the
procedure described above. The general problem of
building a convex hull of $M$ points in $n$ dimensions is
well-known in computational geometry; we employ
the algorithm \verb+quickhull+~\cite{QHull} that
has complexity $\mathcal{O}\left(M^{\lfloor d/2\rfloor}\right)$.
It is sufficient when the number of lines is not too large;
for example, finding 11 expansion regions of a 4-loop integral
with 10 lines takes about 10 seconds on a laptop PC.

The program has been checked against some non-trivial examples 
discussed in ~\cite{Smirnov:2002pj} and ~\cite{Pak:2008cp}.
The code can be downloaded from
\verb+http://www-ttp.particle.uni-karlsruhe.de/~asmirnov/Tools-Regions.htm+

In order to run the program, one has to install the open-source 
package QHull~\cite{QHull}. If the executable
is not in the current directory, \verb+Options[QHull]+ must be updated in 
the file asy.m. The program is loaded with command \verb+<<asy.m+.
The main function is \verb+AlphaRepExpand[ks,ds,cs,hi]+,
where
\verb+ks+ is the list of loop momenta
  (e.g., \verb|{v1,v2}|),
\verb+ds+ are the denominators
  (e.g., \verb|{(p1-v1-v2)^2-M^2,(p1-v2)^2-M^2,(p2+v1+v2)^2-m^2,(p2+v2)^2-m^2,v1^2,v2^2}|),
\verb+cs+ contains the kinematic constraints
  (e.g., \verb|{p1^2->M^2,p2^2->m^2,p1*p2->-S/2}|), and
\verb+hi+ represents the scalings of kinematic invariants with respect to
  the small parameter \verb+x+
  (e.g., \verb|{M->x^0,S->x^0,m->x^1}|).

The output is a list of vectors specifying the scales of the alpha-parameters
factors, or the non-zero components of vectors $\vec{v}_i$.
For the last example above, the output is
\verb+{{0,-2,-2,-2,-2,-4},{0,0,-2,-2,-2,-2},{0,0,-2,0,0,0},{0,0,0,-2,0,2},{0,0,0,0,0,0}}+,
corresponding to the regions (us-1c), (1c-h), (h-1c), (1c-1c), and (h-h).
These regions can be understood by analogy to the example of
Section~\ref{sec:lte}.

\section{Conclusion}

We present an algorithm to find the relevant regions of expansion
for a Feynman integral in a given limit of momenta and masses.
The algorithm is implemented in \verb+Wolfram+ \verb+Mathematica+
language and uses open-source package \verb+QHull+. The
program and examples can be downloaded from our web-page.
In the future, we plan to extend the code in order
to apply it to more general parametric integrals.

{\itshape Acknowledgements}. We would like to thank V.A.Smirnov for
suggesting the topic and constant help, and M. Steinhauser
for the very useful comments.

\bibliography{asy}

\end{document}